\documentclass[a4paper,aps,pra,reprint,superscriptaddress,final]{revtex4-1}
\pdfoutput=1
\usepackage[utf8]{inputenc}
\usepackage[colorlinks,citecolor=blue,linktocpage,breaklinks,hypertexnames=false,pdfpagelabels,draft=false]{hyperref}
\usepackage{microtype}
\usepackage{graphicx}

\usepackage[obeyDraft]{todonotes}
\usepackage[intlimits]{amsmath}
\usepackage{amssymb, amsfonts, amsthm}
\usepackage{enumerate}
\usepackage{quantum}

\theoremstyle{plain}
\newtheorem{lemma}{Lemma}
\newtheorem{thm}[lemma]{Theorem}

\newcommand{\norm}[1]{\left\lVert#1\right\rVert}
\newcommand{\du}[1]{#1^\dagger}
\newcommand{\de}[1]{\ensuremath{\operatorname{d}\!{#1}}}
\DeclareMathOperator{\poly}{poly}
\DeclareMathOperator{\dist}{dist}
\DeclareMathOperator{\trace}{Tr}
\renewcommand{\epsilon}{\varepsilon}

\begin{document}

\title{Rapid mixing and stability of quantum dissipative systems}

\author{Angelo  Lucia}
\affiliation{Dpto. An\'alisis Matem\'atico, Universidad Complutense de Madrid, 28040 Madrid, Spain}

\author{Toby S. Cubitt}
\affiliation{DAMTP, University of Cambridge, Cambridge CB3 0WA, United Kingdom}

\author{Spyridon Michalakis}
\affiliation{Institute for Quantum Information and Matter, Caltech, Pasadena, CA 91125, USA}

\author{David \surname{Pérez-García}}
\affiliation{Dpto. An\'alisis Matem\'atico, Universidad Complutense de Madrid, 28040 Madrid, Spain}

\begin{abstract}
The physics of many materials is modeled by quantum many-body systems with local interactions. If the model of the system is sensitive to noise from the environment, or small perturbations to the original interactions, it will not properly model the robustness of the real physical system it aims to describe, or be useful when engineering novel systems for quantum information processing. We show that local observables and correlation functions of local Liouvillians are stable to local perturbations if the dynamics is rapidly mixing and has a unique fixed point. No other condition is required.
\end{abstract}

\maketitle

Traditionally, the study of quantum many-body systems has focused on constructing simplified models that capture the underlying physics of real materials in order to explain their physical properties and behavior. More recently, quantum information theory has added a complementary perspective, by asking how quantum many-body systems can be artificially engineered to produce useful behavior, such as long-term storage of information~\cite{ToricCode4D,Haah11,RMP.82.1041,Maurer08062012}, or processing of information in a quantum computer \cite{Kitaev20032,RMP.80.1083,RMP.82.1209,briegel2009measurement,Farhi20042001}. This has come full circle, with one of the most important applications of quantum information processing being the simulation of other quantum systems which are computationally intractable by classical means \cite{bloch2012quantum,blatt2012quantum,aspuru2012photonic,houck2012chip}
Whether studying theoretical models of many-body physics, or artificially engineering their dynamics for information processing purposes, it is crucial that the properties of the model are stable under perturbations \emph{to the model itself}. If the physical predictions of a model undergo dramatic changes when the local interactions are modified by a small amount, it is difficult to argue that the idealized model captures the correct physics of the real physical system. Similarly, if the correct behavior of an engineered quantum system relies on infinitely precise control of all the local interactions, the proposal will not be of much practical use.

In the case of closed systems modeled by Hamiltonian dynamics, recent breakthroughs have given rigorous mathematical justification for our intuition that the physical properties of many-body Hamiltonians are stable to small perturbations. Starting with~\cite{Klich:2010,Bravyi-Hastings-Spiros}, it culminated in the work of~\cite{Spiros11} which showed that, under a set of mathematically well-defined and physically reasonable conditions, the properties of gapped many-body Hamiltonians are stable under perturbations to the local interactions.

However, even the most carefully isolated physical systems are susceptible to external noise and dissipation. Broadly, many-body theory has traditionally viewed dissipation as a source of errors to be modeled theoretically and minimized experimentally. Recently, the quantum information ``engineering'' approach has been extended to dissipative quantum systems, with the aim of \emph{exploiting} dissipation. Both theoretical~\cite{Kraus08,verstraete09} and experimental~\cite{trimborn2008mean,witthaut2008dissipation,kordas2012dissipation,PRL.107.080503,2010NatPh6.943B} work has shown that creating many-body quantum states as fixed points of engineered, dissipative Markovian dynamics can be more robust against undesirable noise, both in maintaining coherence of quantum information for longer times~\cite{Kraus08,verstraete09,2011PhRvA..83a2304P}, and in carrying out universal quantum computation via dissipative dynamics~\cite{verstraete09}.

Intuitively, there is an inherent robustness in such proposals: since a dissipative system converges to its steady state eventually, regardless of the state in which it was initialized, the long-term behavior of the system is insensitive to the system's current state. Indeed, this remains the case even if some external process completely changes the state of the system part way through the evolution; if the dissipation is engineered perfectly, the system will inexorably be driven back towards the desired steady state. However, once again, this robustness relies on the hitherto unproven assumption that the physical behavior of the system is insensitive to small implementation errors in engineering the local interactions of the system itself. Therefore, both in justifying theoretical models of real, noisy physical many-body systems, and in the new proposals for exploiting dissipation to carry out quantum information processing tasks, it is crucial to go beyond stability of closed systems, and derive stability results for open, dissipative systems. While earlier works, such as \cite{2011PhRvA..83a2304P},
have produced numerical evidence for stability of particular models, we are interested in producing general analytical results.

In this article, we prove that \emph{rapid mixing implies stability against local perturbations}. Our result shows that rapidly-mixing systems with unique fixed point are stable in the strongest possible sense: all local observables and correlation functions are stable against local perturbations, independent of the system size. This is true not only in the infinite time limit (i.e.\ for the steady state), but also for all intermediate times. In other words, we prove that local observables of the perturbed system are good approximations to the unperturbed observables \emph{throughout the entire evolution}. We prove our result for the more general and difficult case of quantum dissipative Markovian dynamics. 

Single site noise processes, and all ``non-interacting'' dissipative processes trivially satisfy our rapid mixing condition. For interacting models proving estimates on the mixing times is generally a hard task; nonetheless, it is know that dissipative state preparation for graph states (a resource for some error-correcting codes and some quantum computation models) is rapidly mixing \cite{Kastoryano12,verstraete09}. 
Moreover, as classical Markovian dissipative dynamics is a special case of quantum dissipative dynamics, our results also apply to the classical setting; indeed, our results imply stability of classical systems even to quantum perturbations. As an example, we apply our result to prove stability of the important and widely-studied classical Glauber dynamics.

For the sake of simplicity of the exposition, we restrict our attention to translationally-invariant, nearest-neighbor, dissipative interactions on spins arranged on a $D$-dimensional square lattice with periodic boundary conditions.
The proof in the general case follows the same ideas, but becomes notationally involved. It is available in \cite{ourselves}.

\paragraph{Terminology:}
\emph{Rapid mixing} corresponds to the assumption that the convergence of the density matrix $\rho(t)$ of the system to its steady state $\rho_\infty$, as a function of time $t$, is of the form $\norm{\rho(t) - \rho_\infty}_1 \leq c \poly(L) e^{-\gamma t}$ for some constants $c,\gamma$ independent of system size, where $L$ is the linear size of the system.
Since we are considering finite dimensional systems, the exponential convergence with respect to time is a general property;
the non-trivial content of the rapid mixing condition is how $\gamma$ and the multiplicative pre-factor depend on $L$.

\emph{Local perturbation} means that the local interactions of the system can be modified \emph{everywhere}. Indeed, our result applies more generally to arbitrary perturbations composed of a sum of local terms, not only to modifications of the strength of the original local interactions of the system. This is the natural (and standard) model of perturbations in physical systems with local interactions. Note that the total perturbation is a sum of all the local terms, and therefore may diverge with system size regardless of how weak the local perturbations are. Standard perturbation theory breaks down completely in this setting, as the overall perturbation is usually unbounded. It is, instead, the \emph{local} structure of the perturbation that permits stability in our setting.
Moreover, recall that a linear map from operators to operators is called a {\it superoperator}. The \emph{support} of a superoperator is defined to be the smallest set $\Gamma \subset \ZZ_L^D$, such that the operator acts trivially outside of $\Gamma$.

The restriction to local observables and correlation functions, apart from being justified by practical considerations of what can be measured in experiments, also has a fundamental theoretical justification: global observables on the full system cannot be stable to local perturbations. (This is equally true for Hamiltonian systems.) It is easy to construct simple examples that demonstrate this~\footnote{Consider as an example two dissipative processes acting on a qubit and having as a fixed point two pure states with an overlap of $1-\epsilon<1$. Then, by taking the tensor product of $N$ of each process, we note that the fixed points of the composed processes become orthogonal as $N\rightarrow\infty$, for any fixed $\epsilon > 0$.}. But it is also intuitively obvious from the above discussion: global observables can ``see'' the effect of the local perturbations integrated over the entire system, and this effect diverges with system size.

While our result is motivated by the work of~\cite{Spiros11} for Hamiltonian systems, both the result itself and several of the concepts and techniques required for the proof are different in the dissipative case. In the Hamiltonian case, stability is proven under the assumption that the system is frustration-free, has local topological quantum order (LTQO), and is locally-gapped. In the dissipative case, our result derives stability for all rapidly-mixing systems (which can be viewed as the dissipative analogue of the local-gap condition for Hamiltonians), without any need for frustration-freeness (i.e.\ detailed balance), or LTQO for the steady state. We are able to derive the necessary properties of the steady state from the rapid mixing condition alone. Moreover, the technical proof in the Hamiltonian setting relies on the fact that Hamiltonian dynamics is reversible. This is by definition false for dissipative systems, necessitating a different mathematical approach.

\paragraph{Main result.}
Let $\Lambda \simeq \ZZ_L^D$ denote the $D$-dimensional square lattice. The dynamics is then generated by a local Liouvillian $\mathcal L = \sum_{u \in \Lambda} L_u$ (the dissipative analogue of a local Hamiltonian), where each $L_u$ has the well-known Lindblad form (the most general form that preserves complete positivity of the density matrix):
$
  L_u(A) = i [H_u,A] + \sum_j \left[\du{K_{u,j}} A K_{u,j}  
  - \frac12 \left( \du{K_{u,j}}K_{u,j} A + A \du{K_{u,j}}K_{u,j} \right)\right],
$
where $K_{u,j}$ are arbitrary operators and $H_u$ is Hermitian. The $L_u$ terms are related by translation, with each term acting only on $u$ and its neighbors. The evolution of an observable $A$ in the Heisenberg picture is then given by $A(t) = e^{t \mathcal L}(A)$, which is the solution to the differential Liouville master equation $\dot A(t) = \mathcal L A(t)$. We can assume without loss of generality that the strength of the local interactions $L_u$ is bounded as follows (in the completely-bounded norm):
$
  \sup_u \norm{L_u}_{cb}  := \sup_u \sup_n \norm{L_u \otimes \identity_n} \le 1.
$
We will also assume that $\mathcal L$ has a unique fixed point, i.e.\ in the Heisenberg picture $A(\infty) := \lim_{t \to \infty} A(t) = \trace (A \rho_\infty) \identity$ for any observable $A$ \footnote{It is easy to see that in a
system with degenerate fixed points space, a small local perturbation could destroy the degeneracy by selecting some of the fixed points over the others. Therefore it is not possible to generalize our result to such scenario without either requiring extra assumptions or weakening the notion of stability}.

In the Heisenberg picture, the rapid mixing condition states that, for any observable $A$, $A(t)$ converges fast to $A(\infty)$. More precisely, there exist positive constants $c$, $\delta$ and $\gamma$ independent of system size, such that
\begin{equation}\label{eq:rapid-mixing}
  \norm{A(t) - A(\infty)} \le c L^\delta e^{-\gamma t}.
\end{equation}

The perturbed evolution is given by a different local Liouvillian, $\tilde{\mathcal L}$, such that $\tilde{\mathcal L} = \mathcal L + \sum_u E_u$, where the perturbation terms $E_u$ are local, and their strength is bounded by $\epsilon$ (i.e.\ $ \sup_u \norm{E_u}_{cb} \le \epsilon $).
\begin{thm}
  \label{thm:stability}
  For an observable $A$ supported on $X \subset \Lambda$, let $A(t) = e^{t \mathcal L}(A)$ and $\tilde A(t) = e^{t \tilde{\mathcal L}} (A)$ be the time-evolution of the observables in the Heisenberg picture, under the original and perturbed Liouvillians, respectively.

Then for all $t \ge 0$
  \begin{equation}\label{eq:stability}
    \norm{A(t) - \tilde{A}(t)} \le C_X \norm{A} \epsilon,
  \end{equation}
  for some $C_X>0$ not depending on the system size and independent of $t$.
\end{thm}
Note that we do not require the support $X$ of the observable $A$ to be connected. Our result therefore immediately applies to two-point (or, more generally, $k$-point) correlation functions.

In fact, the same result applies to systems with quasi-local interactions, where the interactions $L_u$ and perturbations $E_u$ act on arbitrarily distant spins, but the interaction strength decays exponentially with distance. The results also generalize to interactions with polynomially-decaying strength, and to the non-translationally-invariant case with arbitrary boundary conditions (under natural uniformity conditions that make the concept of scaling with system size meaningful). Moreover, the rapid mixing condition given in Eq.~\eqref{eq:rapid-mixing} can be weakened to slower-than-exponential decay: $\norm{A(t) - A(\infty)} \le c\, L^{\delta} \gamma(t)$, where $\gamma(t)$ is decaying at least as $(1+t)^{-(D+2+\delta+\eta)}$, for some arbitrarily small $\eta > 0$ ~\cite{ourselves}.

\paragraph{Sketch of the proof:}
The main technical tool we need is the Lieb-Robinson bound \cite{Poulin10,Nachtergaele12}. In many-body quantum systems, where the evolution is generated by local interactions, there exists an effective ``light-cone'' outside of which the amount of information that can escape is negligible. The effective velocity that limits the light-cone is called the Lieb-Robinson velocity, and is in general many orders of magnitude smaller than the actual speed of light \footnote{Indeed, the models we consider are non-relativistic, so the existence of such a limit speed is not at all trivial. It could more accurately be described as a speed of sound, as it depends on the microscopic structure of the ``medium'', i.e.\ the local interaction terms.}.

\begin{figure}[h]
  \begin{center}
    \includegraphics[width=240px]{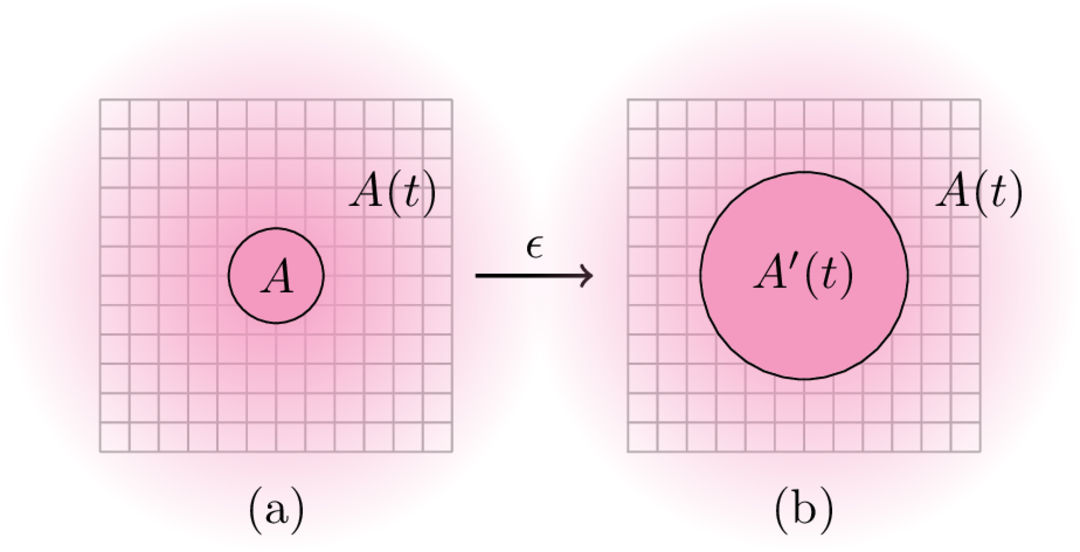}
  \end{center}
  \vspace{-2em}
  \caption{
  \label{fig:Lieb-Robinson}(a) The support of a local observable $A$ spreads linearly in time at the Lieb-Robinson velocity, up to an exponentially small error. (b) The time-evolved observable $A(t)$ can be approximated to small error $\epsilon$ by a local observable $A'(t)$. (Color online)}
\end{figure}

The existence of such light-cones implies that localized observables spread linearly in time, up to negligible tails outside the cones (FIG.~\ref{fig:Lieb-Robinson}). Since the system is rapidly-mixing by assumption, by the time the system has relaxed and reached its steady state, any finite region of the lattice has only had time to interact with a bounded region around it, namely a region of size proportional to the mixing time. There is effectively no further evolution after that time-scale.
This implies that a local observable feels the effects of only part of the total perturbation: the local perturbations acting near the support of the observable. One might then be tempted to consider just this effective perturbation and obtain a bound for the evolution of the observable under examination. However, this is not yet sufficient for our purposes, as this reduced perturbation still scales (sub-linearly) with the system size, so diverges for large system sizes.

We improve on this idea by showing that, under the same conditions, evolution of a local observable can be approximated in a finite region around its support, with a localized evolution that only takes a \emph{finite} time to reach its steady state. Since we are working with a translation invariant model with periodic boundary conditions, the localized evolution we choose is the one given by the global Liouvillian, but defined on a smaller lattice size. After proving this stronger property, it is then straightforward to apply the original approach of restricting the perturbation to a finite region, leading to the proof of the main result.

\paragraph{Proof of main result:}
To fix notation, we will consider a normalized observable $A$ supported on a region $X$, and will denote by $X(s)$ the region $X$ ``grown'' by $s$, i.e.\ $X(s) = \{ u \in \Lambda : \dist(u,X) \le s \}$. Without loss of generality, we can assume that $X(s)$ is always a disjoint union of convex regions~\footnote{When $s$ is sufficiently large for disconnected regions in $X$ to touch each other, we simply merge them by considering their convex hull.}. We will consider the Liouvillian $\mathcal L_s$ acting on $X(s)$, defined by translational invariance and periodic boundary conditions. The evolution of $A$ under this new Liouvillian will be denoted by $A_s(t)$.
Now, since $\mathcal{L}_s$ is none other than the same local Liouvillian on a smaller lattice, the rapid mixing condition of \eqref{eq:rapid-mixing} applies, immediately giving:
\begin{equation}\label{eq:grm}
  \norm{A_s(t) - A_s(\infty)} \le c_X (1+s)^\delta \gamma(t),
\end{equation}
for some appropriate constant $c_X$, recalling that the linear size of $X(s)$ is bounded by $\mathrm{diam}(X)+2s$.

Consider a superoperator $\mathcal T$ supported on a region $Y$, such that $d=\dist(X,Y) >0$, and assume that $\mathcal T(\identity) = 0$. The dissipative version of the Lieb-Robinson bound states that there exists some positive constants $k_X$, $v$ and $\mu$, independent of system size, such that for all $t \ge 0$:
$
 \norm{\mathcal T(A(t))} \le k_X \norm{\mathcal T}_{cb} (e^{v t} - 1) e^{-\mu d} .
$
A known consequence of Lieb-Robinson bounds is that we can approximate the evolution of a local observable by a \emph{localized} evolution, i.e.\ by a time-evolved observable whose support only grows linearly with time. Since Lieb-Robinson bounds depend only on the microscopic structure of the evolution, the presence of a boundary condition has a negligible effect on the localized evolution of local observables. Therefore, one may add periodic boundary conditions to the localized evolution coming from the standard Lieb-Robinson bounds, while still obtaining a good approximation for the original evolution of the local observables. More formally,
we obtain the following bound, valid for all $s \ge 0$:
\begin{equation}\label{eq:lrb}
  \norm{A(t) - A_s(t)} \le k_X (e^{vt} - 1) e^{-\mu s}.
\end{equation}

A number of properties of the system can be derived from equations \eqref{eq:grm} and \eqref{eq:lrb}. By the definition of the fixed point, we have that $A(\infty) = \trace (A \rho_\infty)\identity = \trace (A(t) \rho_\infty)\identity$. Then by the triangle inequality:
 \begin{multline*}
  \norm{A(\infty) - A_s(\infty)}
  = \abs{ \trace[A\rho_\infty] - \trace [A \rho^s_\infty ]} \\
  \le \abs{ \trace[(A(t) - A_s(t)) \rho_\infty]}
     + \abs{ \trace[(A_s(t) - A_s(\infty)) \rho_\infty] }\\
  \le \norm{A(t) - A_s(t)} + \norm{A_s(t) - A_s(\infty)}.
 \end{multline*}
Together with eqs.~\eqref{eq:lrb} and~\eqref{eq:grm} and choosing $t$ linear in $s$, it implies that $\norm{A(\infty) - A_s(\infty)}$ decays with $s$. This in turn implies a stronger convergence bound for $A(t)$, since
\begin{multline}
	\label{eq:lrm}
    \norm{A(t) - A(\infty)} \le \norm{A(t) - A_s(\infty)} + \norm{A_s(\infty) - A(\infty)}  \\
    \le 2 \norm{A(t) - A_s(t)} + 2 \norm{A_s(t) - A_s(\infty)} .
\end{multline}
Again by applying eq.~\eqref{eq:lrb} and \eqref{eq:grm}, the r.h.s.\ is bounded by a decaying function $\Delta(t)$, if $s$ is chosen to scale linearly in $t$. The big difference with respect to the rapid mixing condition is that we have managed to remove the dependence on the system size from the pre-factor of the r.h.s., since the bounds in eq.~\eqref{eq:lrb} and \eqref{eq:grm} are system size independent. Of course, this was possible because $A$ is a local observable. Note that the assumption made on $\gamma(t)$ implies that $\Delta(t)$ goes to zero at least as $(1+t)^{-(D+2+\eta)}$.

Once we have established such size-independent bounds, we can directly show -- by one last application of Lieb-Robinson bounds -- stability of the evolution of $A(t)$. Let us decompose the quantity we want to bound as follows
$A(t) - \tilde A(t) = \sum_u \int_0^t e^{(t-s) \tilde{\mathcal L}} E_u A(t) \de s .$
Let us take norms and use the fact that $e^{(t-s) \tilde{\mathcal L}}$ is norm-contractive:
\begin{equation}\label{eq:decomposition}
  \norm{A(t) - \tilde A(t)} \le \sum_u \int_0^t \norm{E_u A(t)} \de s.
\end{equation}
For each $u\in \Lambda$, call $d = \dist(X,u)$ and fix a time scale $t_0 = t_0(d)$ to be determined later.
For short times, i.e. for times $t \le t_0$, we can apply the standard Lieb-Robinson bounds and thus
$
  \int_0^{t_0} \norm{E_u A(t)} \de s \le k_X\, \epsilon\,  e^{v t_0 - \mu d}.
$
For long times, i.e. $t \ge t_0$, we bound the integral by using eq.~\eqref{eq:lrm} and the fact that $E_u(\id) = 0$~\footnote{Since both $L_u$ and $L_u + E_u$ are of the Lindblad form, we have that $L_u(\id) = 0$ and that
$0 = E_u(\id) + L_u(\id) = E_u(\id)$}:
$
    \int_{t_0}^\infty \norm{E_u A(t)} \de s
    \le \epsilon \int_0^{t_0} \Delta(s) \de s
$.
We can now choose $t_0(d) = \tfrac{\mu}{2v} d$, such that the integral is entirely bounded by a function decaying in $d$. By putting this back into eq.~\eqref{eq:decomposition}, we can sum over all terms $u$ and obtain the claimed result:
\begin{equation*}
  \norm{A(t) - \tilde A(t)}
  \le \epsilon \sum_u \big( k_Xe^{-\frac{\mu}{2}d}
    + \int_0^{\mu d/2v }\Delta(s) \de s \big)
  \le C_X\, \epsilon.
\end{equation*}
The sum is convergent because $\int_0^d \Delta(s) \de s$ decays to zero at least as fast as $(1+t)^{-(D+1+\eta)}$.

\paragraph{Glauber dynamics:}
One of the systems which satisfies the conditions of our theorem is classical Glauber dynamics \cite{Martinelli97} (the continuous-time version of the Metropolis algorithm), in the regime in which it has a system size independent Log-Sobolev constant. By embedding this dynamics into a quantum Liouvillian in a careful way, our result immediately implies that Glauber dynamics is stable against local perturbations (even those that do not preserve detailed balance). To the best of our knowledge, this result is new even to the classical literature. (Related results, but with different assumptions, were given in \cite{MR814713}.) Given the importance of Glauber dynamics to sampling from the thermal distributions of classical spin systems \cite{Liggett85,Martinelli97}, we expect our results to have applications also to classical statistical mechanics.

\paragraph{Conclusions:}
We have considered the influence that a small but extensive perturbation to the generators of a dissipative quantum many-body master equation can have on the evolution of local observables. We have shown that, if the system relaxes to its unique fixed point sufficiently fast, the observables are stable to such local perturbations throughout the entire evolution: the effect of the the observables depends linearly on the microscopic strength of the perturbation, independently of the system size, even though the magnitude of the overall perturbation diverges with system size. Stability is therefore a result of the local structure of the perturbations.

While the requirement of rapid mixing does not cover all possible interesting quantum systems, the result already has important applications in well-studied models: it applies to dissipative state preparation of graph states \cite{Kastoryano12}, a resource for universal quantum computation; to classical Glauber dynamics, one of the most important models in statistical mechanics; and to the modeling of local noise -- e.g.\ the physically important case of independent local depolarizing noise -- as well as any other noise model which acts independently on every particle in the system. The latter case justifies the choice of a particular type of noise in a theoretical model without requiring perfect knowledge of the form of physical noise (which is essentially unknowable by definition).

\begin{acknowledgments}
  T.\,S.\,C.\ is supported by the Royal Society, and was previously supported by the Juan de la Cierva program of the Spanish MICIM.
  A.\,L.\ and D.\,P.-G.\ acknowledge support from MINECO (grant MTM2011-26912), Comunidad de Madrid (grant QUITEMAD+-CM, ref. S2013/ICE-2801) and the European CHIST-ERA project CQC (funded partially by MINECO grant PRI-PIMCHI-2011-1071).
  A.\,L.\ is supported by MINECO FPI fellowship BES-2012-052404.
  S.\,M.\ acknowledges funding provided by IQIM, an NSF Physics Frontiers Center with support of the Gordon and Betty Moore Foundation through Grant \#GBMF1250, and AFOSR Grant \#FA8750-12-2-0308.
  The authors would like to thank the hospitality of the Centro de Ciencias Pedro Pascual in Benasque, where part of this work was carried out.
\end{acknowledgments}

\bibliography{bibliography}
\end{document}